\documentclass{article}
\usepackage{amsmath,amscd,amsfonts,amssymb,amsthm}
\usepackage{graphicx}
\usepackage{epstopdf}
\usepackage{a4wide}
\usepackage{graphicx}
\usepackage{subfigure}
\usepackage[latin1]{inputenc}
\usepackage{color}
\usepackage{caption}
\usepackage{url}

\newcommand{\subj}[1]{\par\noindent{\bf Mathematics Subject Classification 2010: }#1.}
\newcommand{\keyw}[1]{\par\noindent{\bf Keywords: }#1.}

\theoremstyle{definition}

\theoremstyle{remark}

\def\a{\alpha}
\def\t{\tau}
\def\p{\psi}

\def\LD{{^CD_{a+}^{\a,\p}}}

\def\LI{{I_{a+}^{\a,\p}}}

\begin{document}

\title{What is the best fractional derivative to fit data?}

\author{Ricardo Almeida\\
{\tt ricardo.almeida@ua.pt}}

\date{Center for Research and Development in Mathematics and Applications (CIDMA)\\
Department of Mathematics, University of Aveiro, 3810--193 Aveiro, Portugal}

\maketitle

\begin{abstract}
The aim of this work is to show, based on concrete data observation, that the choice of the fractional derivative when modelling a problem is relevant for the accuracy of a method. Using the least squares fitting technique, we determine the order of the fractional differential equation that better describes the experimental data, for different types of fractional derivatives.
\end{abstract}
\subj{26A33, 34A08, 90C30}
\keyw{ Fractional calculus, modelling, optimization}

\section{Introduction}

Fractional differential equations (FDE) are an extension of ODE's, where the order of the derivative may take any real positive value. For this reason, very often we can model an experimental dynamic more efficiently by considering the problem formulated as a FDE.
For example, they have found applications in viscoelastic \cite{Zerpa}, mechanics \cite{Carpinteri,Katsikadelis}, economy \cite{Machado}, signal processing \cite{Wang}, biology \cite{Almeida}, logistic population models \cite{Yuzbasi}, etc.
A question that always arises when doing so is what fractional derivative is to be considered, since there are various definitions in the literature \cite{Samko}. Usually, the answer to this question depends on the problem and on the observed data.
One of the most common definition, at least for real world applications, is the Caputo fractional derivative. Given a function $f\in C^n[a,b]$ and a positive real $\a$, let $n\in \mathbb N$ be such that $\a\in(n-1,n)$. The Caputo fractional derivative of $f$ of order $\a$ is given by the integral
$${^CD_{a+}^{\a}} f(x):=\frac{1}{\Gamma(n-\a)}\int_a^x (x-t)^{n-\a-1}f^{(n)}(t)\,dt.$$
Here, $\Gamma$ denotes the Gamma function:
$$\Gamma(x):=\int_0^\infty  t^{x-1} e^{-t}\,dt, \quad x>0.$$
This special function is an extension of the factorial function, in the sense that
$$\Gamma(x+1)=x\Gamma(x), \, x>0 \quad \mbox{and} \quad \Gamma(k)=(k-1)!, \, k\in\mathbb N.$$
One way to overcome this vast number of definitions for fractional derivatives is to consider a more general definition for fractional operators, and then determine which ones fit better with the data \cite{Samko}. Let  $\a>0$, $f\in L^1[a,b]$ and $\p\in C^1[a,b]$ be an increasing function such that $\p'(x)\not=0$, for all $x\in [a,b]$.
The fractional integral of $f$ with respect to $\p$ is defined as
$$\LI f(x):=\frac{1}{\Gamma(\a)}\int_a^x \p'(t)(\p(x)-\p(t))^{\a-1}f(t)\,dt.$$
Fractional derivatives are defined in the following way. If $f,\p\in C^n[a,b]$ (with $\p$ increasing and $\p'(x)\not=0$, for all $x$), the fractional derivative of $f$ with respect to $\p$ is given by \cite{Almeida2}
$$\LD f(x):=\frac{1}{\Gamma(n-\a)}\int_a^x \p'(t)(\p(x)-\p(t))^{n-\a-1}\left(\frac{1}{\p'(t)}\frac{d}{dt}\right)^nf(t)\,dt,$$
where $\a\in(n-1,n)$ and $n\in\mathbb N$. We recall the Mittag--Leffler function $E_\a$: given $\a>0$ and $x\in\mathbb R$, this special function is given by the series
$$E_\a(x) := \sum_{k=0}^\infty \frac{x^k}{\Gamma(\alpha k + 1)}.$$
It can be seen as a generalization of the exponential function, since when $\a=1$, we have
$$E_\a(x)=\exp(x).$$
For example, we have the following:
\begin{align*}\LD (\p(x)-\p(a))^{\beta-1}&=\frac{\Gamma(\beta)}{\Gamma(\beta-\a)}(\p(x)-\p(a))^{\beta-\a-1}, \quad \beta>n, \quad \mbox{and}\\
\LD E_\a(\lambda(\p(x)-\p(a))^\a)& =\lambda E_\a(\lambda(\p(x)-\p(a))^\a), \quad \lambda \in\mathbb R.\end{align*}
Fractional integral and fractional derivative are the inverse operation of each other, in the sense that
$$\LI \LD f(x)=f(x)-\sum_{k=0}^{n-1}\frac{\left(\frac{1}{\p'(x)}\frac{d}{dx}\right)^k f(a)}{k!}(\p(x)-\p(a))^k$$
and
$$\LI\LD f(x)=f(x)-f(a).$$
Based on experimental data, our goal is to test several FDE's and see which ones better describe the problems.
All the numerical computations are done in MAPLE, using the \textit{Statistics} package and the Maple tool \textit{NonlinearFit}. This tool fits a nonlinear model to a given data, by minimizing the least-squares error. Suppose that the $n$ points  $(\tilde x_i,\tilde y_i)$ are the original data and the model has the form $F(A,x)$, where the parameter $A$ is a $m$ dimensional vector. The objective is to find the values of the parameters for the model such that the squared residuals
$$\mbox{Error}= \sum_{i=1}^{n}(\tilde y_i-F(A,\tilde x_i) )^2$$
is a minimum. In each section, we will fit three kinds of models to the data points. First, we consider the classical model, which corresponds to the case  $\a=1$. Secondly, we consider the method to be fit when the process is described by the Caputo fractional derivative, that is, when we take $\p_1(x):=x$. In the third model, we consider three different kernels $\p_2(x):=(x+1)^b$, $\p_3(x):=\ln(x+1)$ and $\p_4(x):=\sin(x/b)$, with $b>0$, and choose the one that better fits with the original data. We test the efficiency of the model, by comparing the error of the classical model, $\mbox{Error}_C$, with the error given by the fractional model, $\mbox{Error}_F$:
$$\mbox{Efficiency}=\frac{\mbox{Error}_C-\mbox{Error}_F}{\mbox{Error}_C}\times 100.$$
We will study four problems: the Gross Domestic Product (Section \ref{Sec:GDP}), Newton's law of cooling (Section \ref{sec:newton}), the Bombay Plague Epidemic (Section \ref{sec:Bombay}), and the World population growth (Section \ref{sec:world}). The problems considered are described by a linear differential equation because when we replace after the ordinary derivative by a fractional derivative, we know the exact solution to the fractional differential equation.

\section{Gross Domestic Product}
\label{Sec:GDP}

\begin{figure}[h]
\centering
\subfigure[GDP in the USA: linear case.]{\includegraphics[width=6cm]{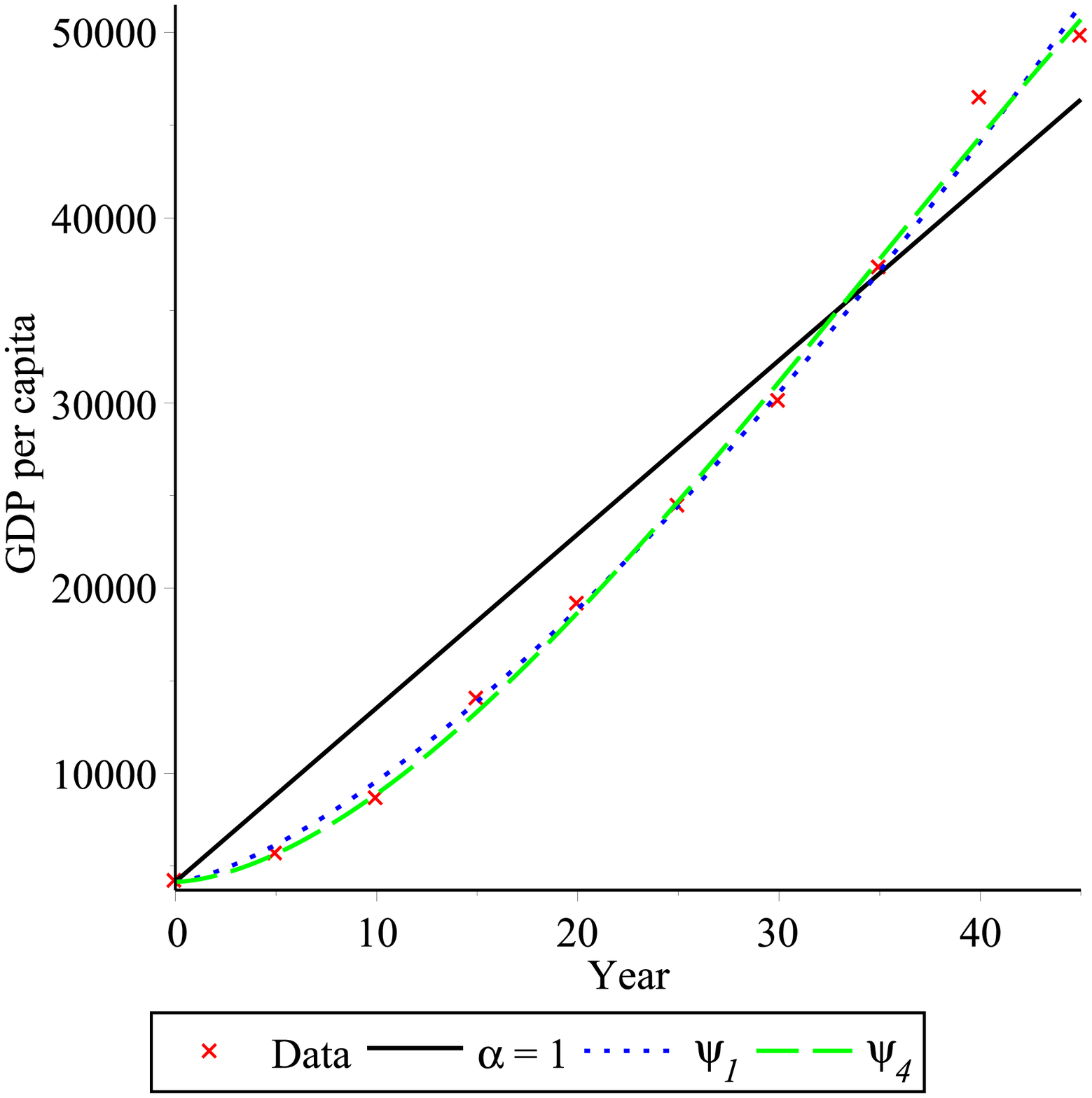}}
\subfigure[GDP in the USA: exponential case.]{\includegraphics[width=6cm]{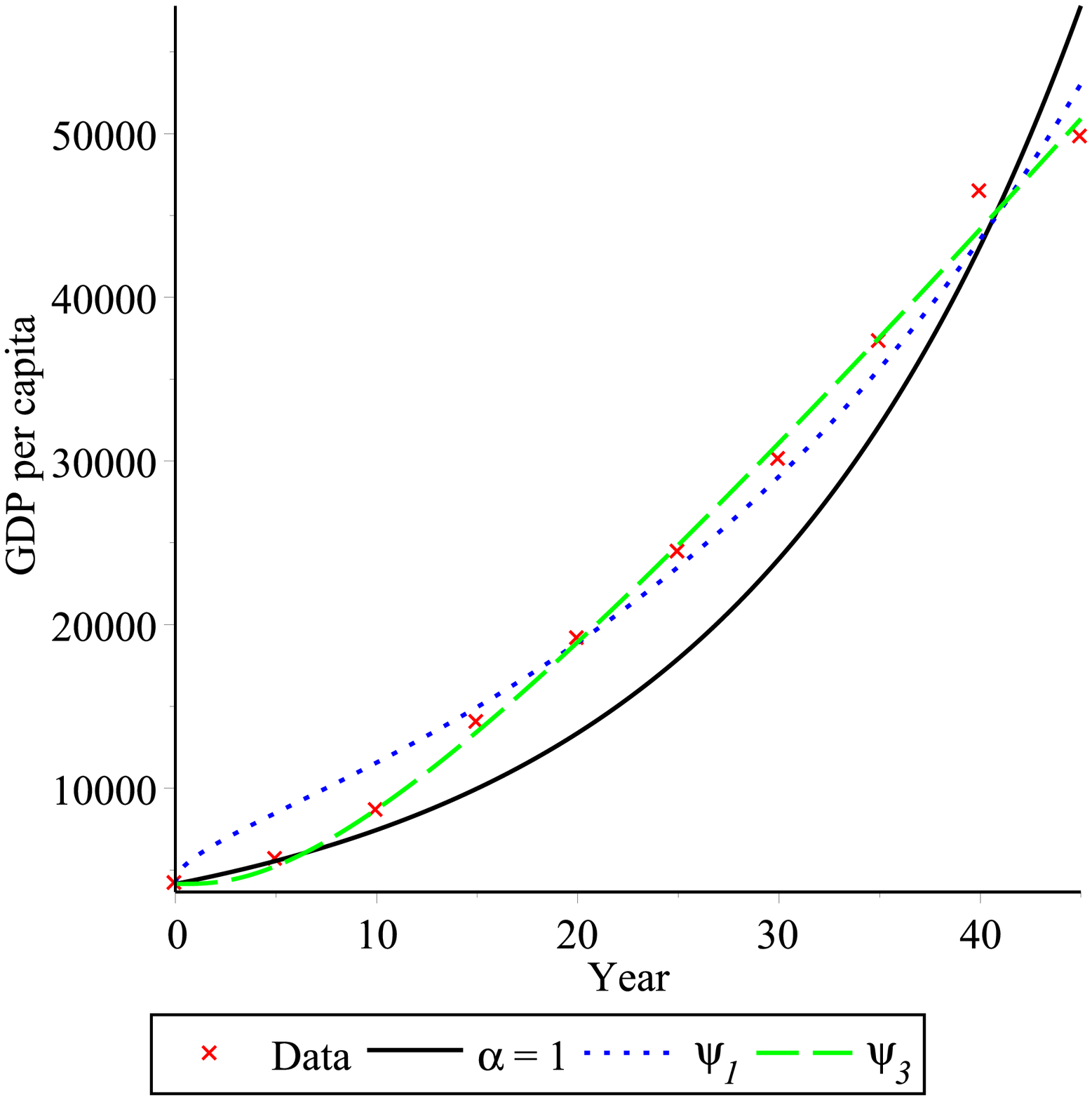}}
\subfigure[GDP in the UK: linear case.]{\includegraphics[width=6cm]{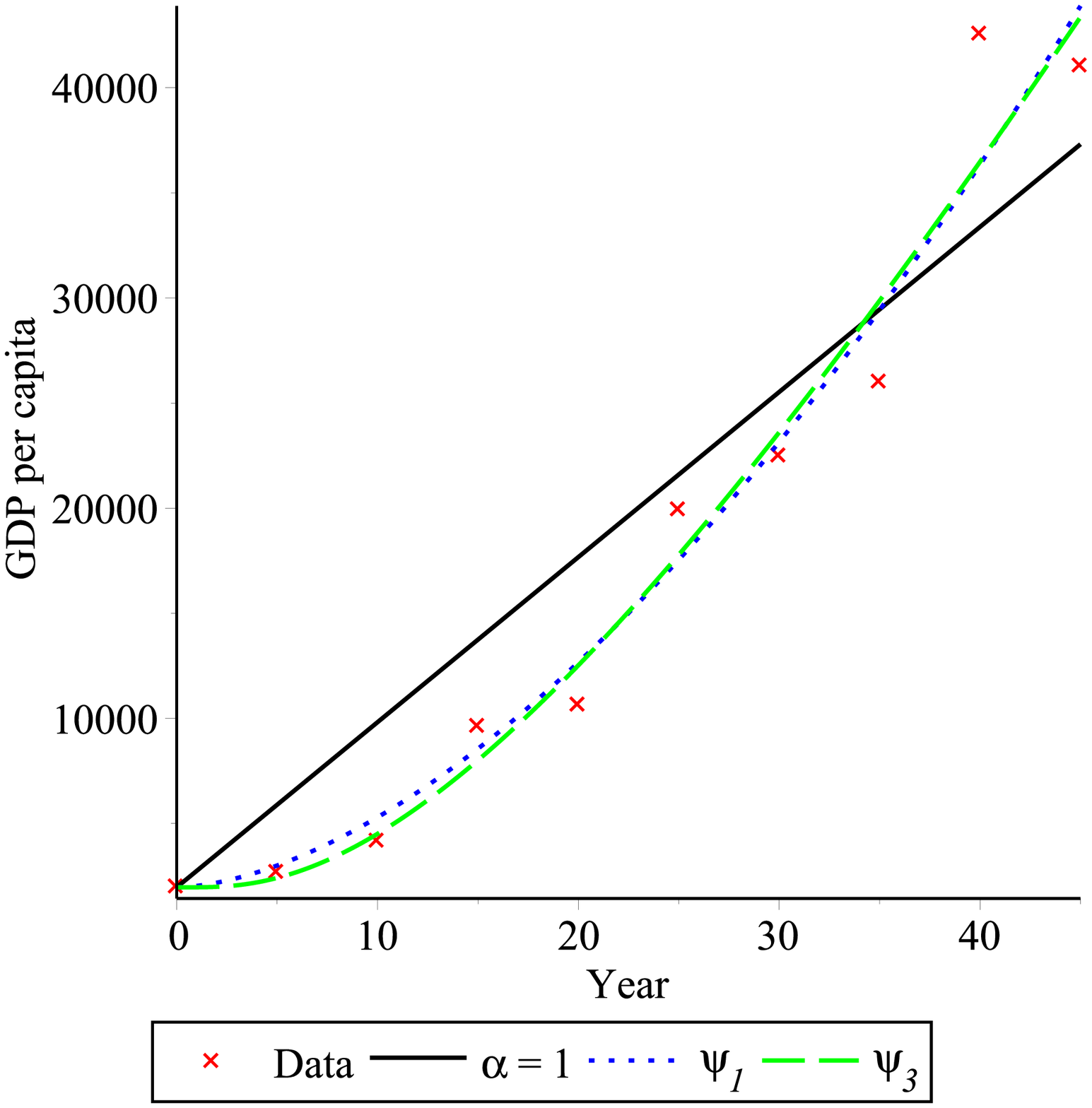}}
\subfigure[GDP in the UK: exponential case.]{\includegraphics[width=6cm]{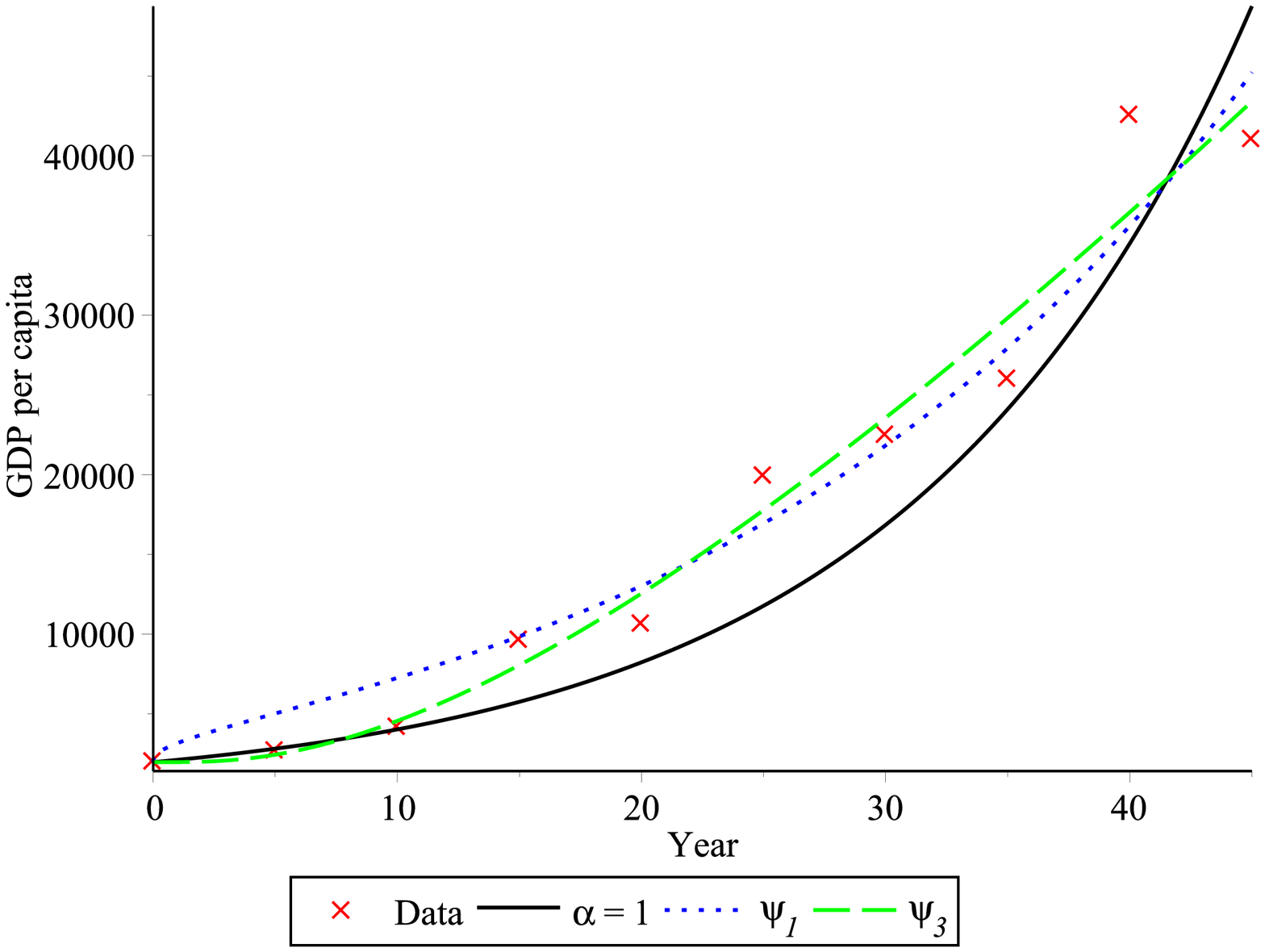}}
\caption{Gross Domestic Product in the USA and UK.}
\label{fig:GDP}
\end{figure}

The Gross Domestic Product (GDP) is one of the most important of all economic statistics, used to measure the performance of a country's economy.  It represents the total of finished goods and services produced  over a specific time period.
We consider in this section the GDP of two countries: the USA and the UK. We will describe  them by a linear and by an exponential model:
$$S'(t)=k \quad \mbox{and} \quad  S'(t)=kS(t).$$
If we consider the model described by the FDE's,
$${^CD_{0+}^{\a,\p}} S(t)=k \quad \mbox{or by} \quad  {^CD_{0+}^{\a,\p}} S(t)=k S(t),$$
with $\a\in(0,1)$, the solutions are given by
$$S(t)= S_0+ {I_{0+}^{\a,\p}} k \quad \mbox{and} \quad S(t)=S_0E_\a(k(\p(t)-\p(0))^\a),$$
respectively.
To get a better accuracy for the model, we will allow the fractional order $\a$ to be arbitrary, and not only on the open interval $(0,1)$.
For the data, we use the ones available from the World Bank  \cite{data:GDP}, which consist in GDP per capita between the years 1966 until 2011, counted every 5 years. The variable $t$, in years, starts at $t=0$, corresponding to the year 1966, and $S(T)$ is the GDP per capita, in US$\$ $.  In Tables
\ref{tab:USA_linear} and \ref{tab:USA_exponential} we present the results obtained when considering the USA case, with linear and exponential growth, and in Tables \ref{tab:UK_linear} and \ref{tab:UK_exponential}, the UK case.
{\small
\begin{center}\begin{tabular}{|c|c|c|c|c|}
\hline
Model & $k$ &  $\a$ & Error & Efficiency\\
\hline
Classical  &  $938.653334$ & $1$ & $1.163673\times10^8$ &\\
\hline
Fractional $\p_1$ & $251.518736$  &$ 1.439934$ &$ 9.895013\times10^6$ &$91.496740\%$ \\
\hline
Fractional $\p_4$ ($b=50$) &$ 1.050515$ &$1.664986$  &$ 7.346859\times10^6$ &$93.686491\%$ \\
\hline
\end{tabular}\captionof{table}{GDP in the USA: linear case.}\label{tab:USA_linear}
\end{center}
\begin{center}\begin{tabular}{|c|c|c|c|c|}
\hline
Model & $k$ &  $\a$ & Error & Efficiency\\
\hline
Classical  &  $ 0.058556$ & $1$ & $2.302460\times10^8$ &\\
\hline
Fractional $\p_1$ & $0.267899$  &$0.400291$ &$4.129788\times10^7$ &$82.063587\%$ \\
\hline
Fractional $\p_3$ &$1.501965$ &$4.880700$  &$8.241729\times10^6$ &$96.420468\%$ \\
\hline
\end{tabular}\captionof{table}{GDP  in the USA: exponential case.}\label{tab:USA_exponential}
\end{center}
\begin{center}\begin{tabular}{|c|c|c|c|c|}
\hline
Model & $k$ &  $\a$ & Error & Efficiency\\
\hline
Classical  &  $785.672510$ & $1$ & $ 2.315927\times10^8$ &\\
\hline
Fractional $\p_1$ & $105.319149$  &$1.683547$ &$ 7.114080\times10^7$ &$69.687165\%$ \\
\hline
Fractional $\p_3$ &$9.274772\times 10^4$ &$5.961180$  &$ 6.947902\times10^7$ &70.408906$\%$ \\
\hline
\end{tabular}\captionof{table}{GDP  in the UK: linear case.}\label{tab:UK_linear}
\end{center}
\begin{center}\begin{tabular}{|c|c|c|c|c|}
\hline
Model & $k$ &  $\a$ & Error & Efficiency\\
\hline
Classical  &  $0.071713$ & $1$ & $2.562247\times10^8$ &\\
\hline
Fractional $\p_1$ & $0.359075$  &$0.33255$ &$9.980382\times10^7$ &$61.048327\%$ \\
\hline
Fractional $\p_3$ &$4.412682$ &$5.873892$  &$6.951255\times10^7$ &$72.870476\%$ \\
\hline
\end{tabular}\captionof{table}{GDP  in the UK: exponential case.}\label{tab:UK_exponential}
\end{center}
}
Either way, the best model to describe the problem is when we consider linear growth. This allows us to predict the GDP in future years more accurately. For example, in the year 2014, the GDP in the USA is about 54629$\$ $ and in the UK 46332$\$ $. Using the classical model and the values of Tables \ref{tab:USA_linear} and \ref{tab:UK_linear}, the predictions are 49202$\$ $ for the USA, and 39672$\$ $ for the UK. So, the errors are
$$54629-49202=5427 \quad \mbox{and} \quad 46332-39672=6660,$$
respectively. When considering the fractional models, the predictions are 54305$\$ $ for the USA and 47616$\$ $ for the UK. The errors are now
$$54629-54305=324 \quad \mbox{and} \quad 47616-46332= 1284,$$
respectively.
In Figure \ref{fig:GDP} we show the optimal curves, for the USA and UK cases, considering the linear and exponential growths.

\section{Newton's law of cooling}
\label{sec:newton}

\begin{figure}[h]
\centering
\includegraphics[width=7cm]{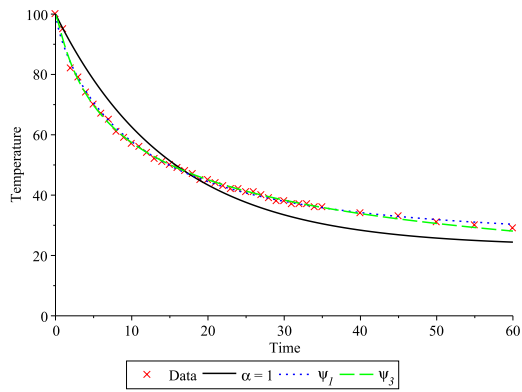}
\caption{Newton's law of cooling.}
\label{fig:newton}
\end{figure}

Newton's law of cooling states that the rate of change of the temperature of a body is directly proportional to the difference between its own temperature and the ambient temperature, provided the difference is small.  This statement leads to the classic equation of exponential decline over time, which can be applied to many phenomena. Newton was able to show that temperature change follows the ODE
$$T'(t)=k(T(t)-T_a),$$
where $T(t)$ is the temperature of the body at a given time $t$, $T_a$ is the ambient temperature, and $k$ is a cooling constant, specific to the object. The solution of this EDO is the exponential function
$$T(t)=T_a+(T_0-T_a)\exp(kt),$$
where $T_0$ is the  starting temperature of the object. If we consider the problem modeled by a FDE of order $\a>0$, then we get that the solution of the equation
$${^CD_{0+}^{\a,\p}} T(t)=k(T(t)-T_a)$$
is the function
$$T(t)=T_a+(T_0-T_a)E_\a(k(\p(t)-\p(0))^\a).$$
To test the accuracy of the model an experiment was carried out. With a  beaker filled with 100ml of water, originally at 100$^o\,C$, and with a thermometers in the beaker, the temperature of the water was measured every minute. The ambient temperature for this investigation was 23$^o\,C$, and the results obtained can be viewed in \cite{newton}. In Table \ref{tab:newton} we present the numerical results for this problem, and in Figure \ref{fig:newton} we compare the plots for the classical model with the two fractional models.
{\small
 \begin{center}\begin{tabular}{|c|c|c|c|c|}
\hline
Model & $k$ &  $\a$ & Error & Efficiency\\
\hline
Classical  & $-0.0665349$  & $1$ & $940.640679$ &\\
\hline
Fractional $\p_1$ & $-0.130333$  &$0.798038$ &$39.844721$ &$95.764087\%$ \\
\hline
Fractional $\p_3$ &$-0.249423$ &$1.428019$  &$26.691909$ &$97.162369\%$ \\
\hline
\end{tabular}\captionof{table}{Newton's law of cooling.}\label{tab:newton}
\end{center}
}

\section{Bombay Plague Epidemic, 1905-6}
\label{sec:Bombay}

\begin{figure}[h]
\centering
\subfigure[Total deaths.]{\includegraphics[width=6cm]{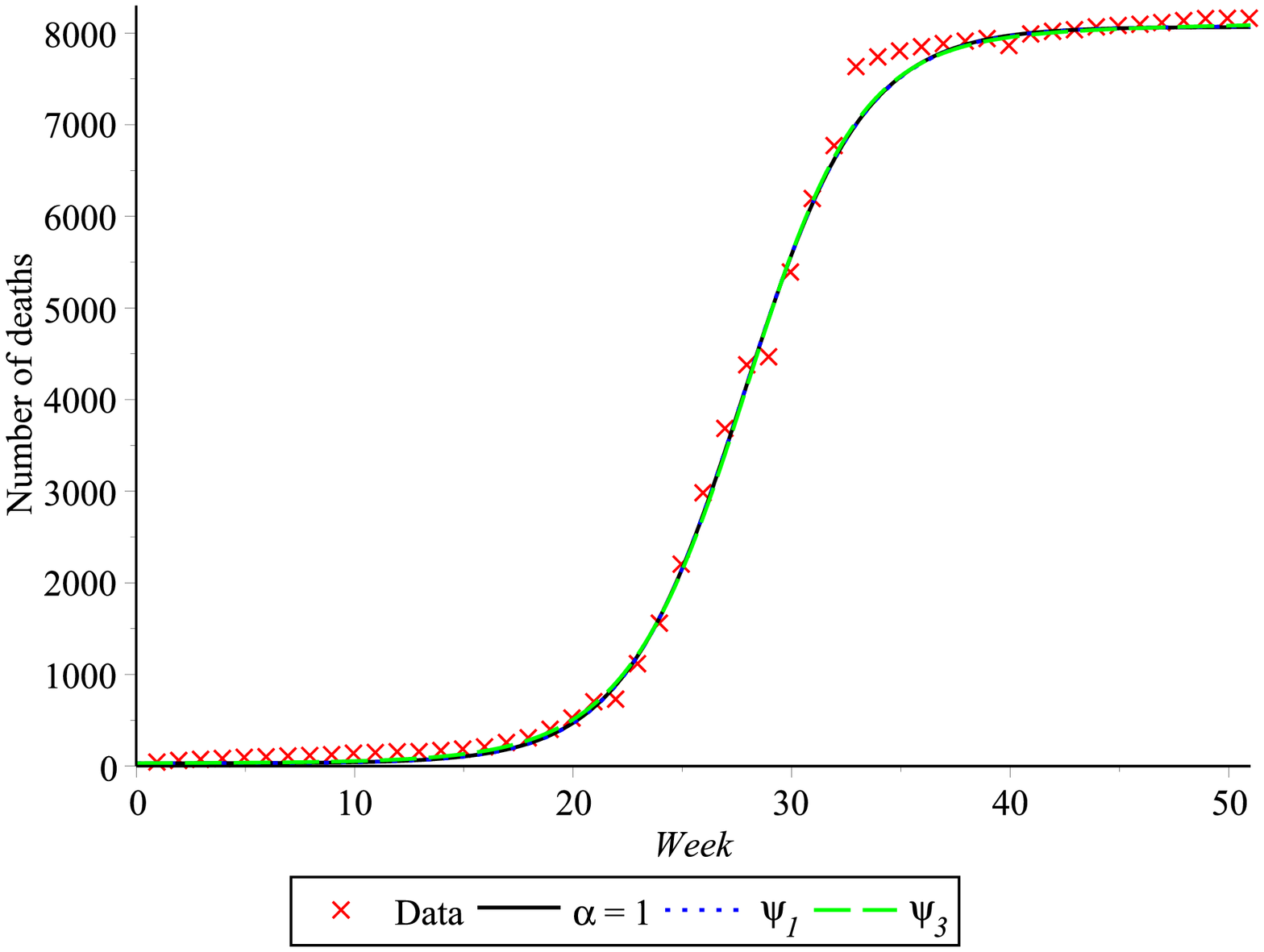}}
\subfigure[Deaths per week.]{\includegraphics[width=6cm]{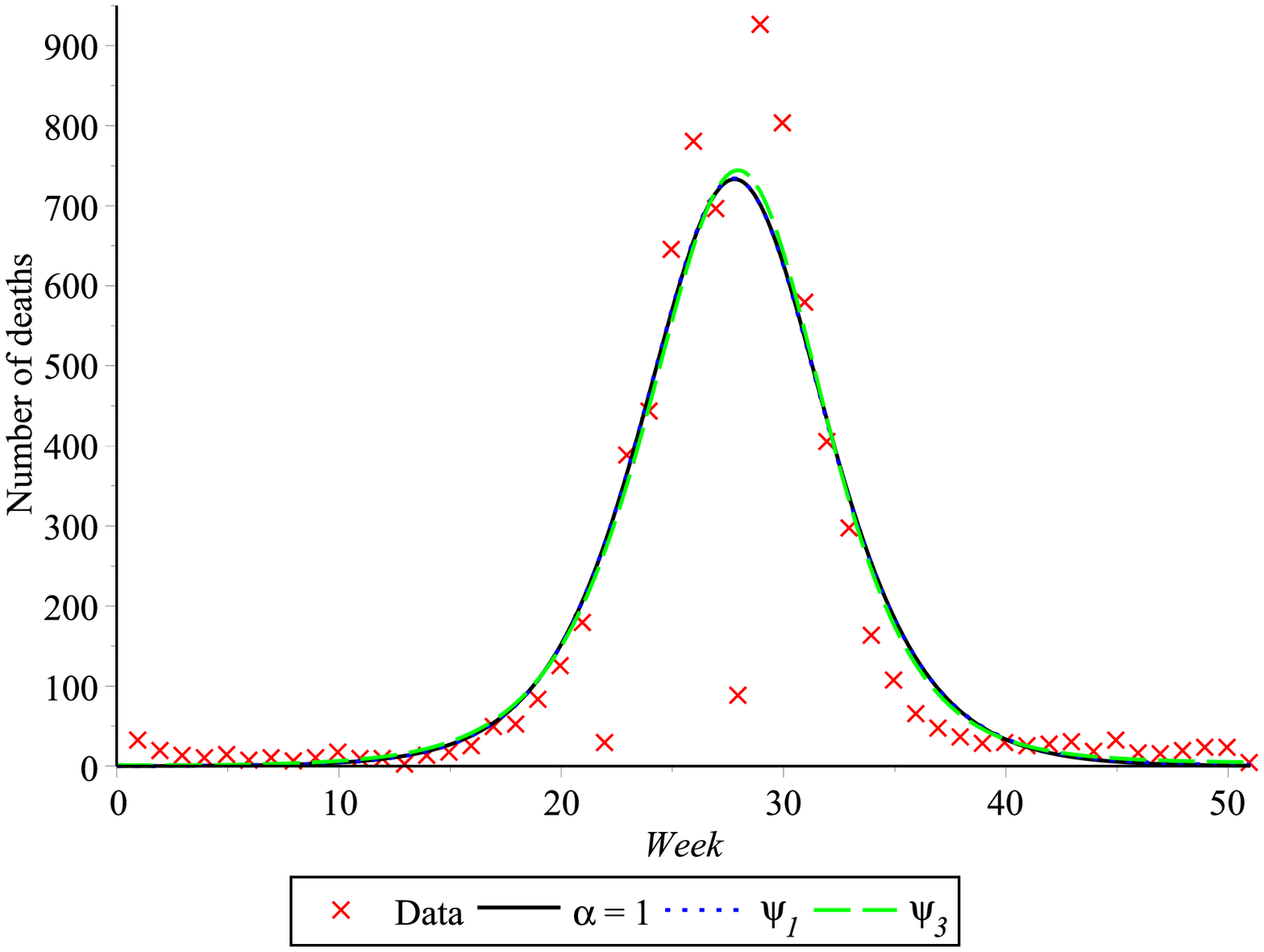}}
\caption{Bombay Plague Epidemic.}
\label{fig:bombay}
\end{figure}

Between 1896 and 1914, India was struck by a series of major epidemics: malaria, cholera, Spanish influenza, and the bubonic plague. In Bombay City alone, the number of deaths were up to 183984 on that period. Kermack and McKendrick proposed a model \cite{Kermack} based on the number of deaths from October 1905 until September 1906 \cite[Table IX]{data:bombay}, counted every week. If $R(t)$ denotes the number of deaths in the week $t$, for $t\in\{0,\ldots,51\}$, with $t=0$ corresponding to 1st of October 1905, the authors proposed the model obtained from the ODE
$$R'(t)=A/\cosh^2(Bt+C).$$
The solution is given by the function
$$R(t)=R_0+\int_0^x \frac{A}{\cosh^2(B\t+C)}d\t.$$
When we consider the fractional model,
$${^CD_{0+}^{\a,\p}} R(t)=A/\cosh^2(Bt+C),$$
with $\a\in(0,1)$, applying the fractional integral to both sides of the equation, we deduce that
\begin{align*}
R(t) & = R_0+{I_{0+}^{\a,\p}}\frac{A}{\cosh^2(Bt+C)} \\
 & =R_0+\frac{1}{\Gamma(\a)}\int_0^x \p'(\t)(\p(t)-\p(\t))^{\a-1}\frac{A}{\cosh^2(B\t+C)}d\t.
\end{align*}
The results obtained are presented in Table  \ref{tab:bombay}. In Figure \ref{fig:bombay} we show the plots for the total number of deaths, and for the number of deaths per week. To obtain a better model we will consider $\a>0$. As we can see, when we model the problem using the Caputo fractional derivative the gain in efficiency is very small. However, considering another kernel we obtain a gain of approximately $5\%$.
{\small
\begin{center}\begin{tabular}{|c|c|c|c|c|c|c|}
\hline
Model & $A$ & $B$ & $C$ & $\a$ & Error & Efficiency\\
\hline
Classical  &  $736.714057$ & $0.183346$ & $-5.070156$ &  $1$ &  $6.309364\times10^5$ &\\
\hline
Fractional $\p_1$ & $ 733.259286$ & $0 .184230$& $-5.088524$ &$ 1.002834$ &$6.302366\times10^5$ &$0.110908\%$ \\
\hline
Fractional $\p_3$ &$2.217880\times10^4$ &$0.194156$  &$-5.478100$ & $1.017019$& $5.983701\times10^5$ &$5.161572\%$ \\
\hline
\end{tabular}\captionof{table}{Bombay Plague Epidemic.}\label{tab:bombay}
\end{center}
}

\section{World population growth}
\label{sec:world}

\begin{figure}[h]
\centering
\subfigure[Africa.]{\includegraphics[width=6cm]{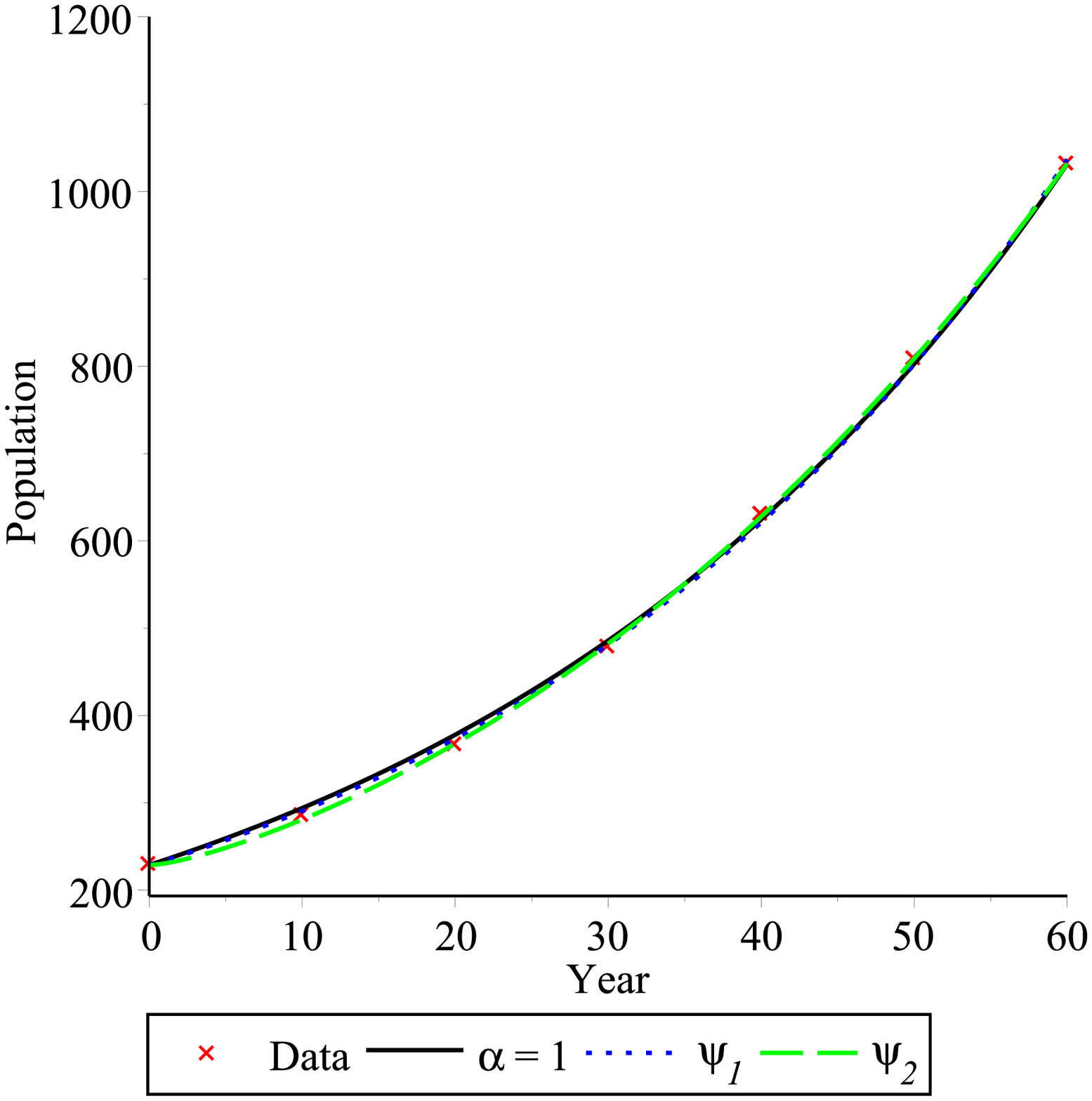}}
\subfigure[America.]{\includegraphics[width=6cm]{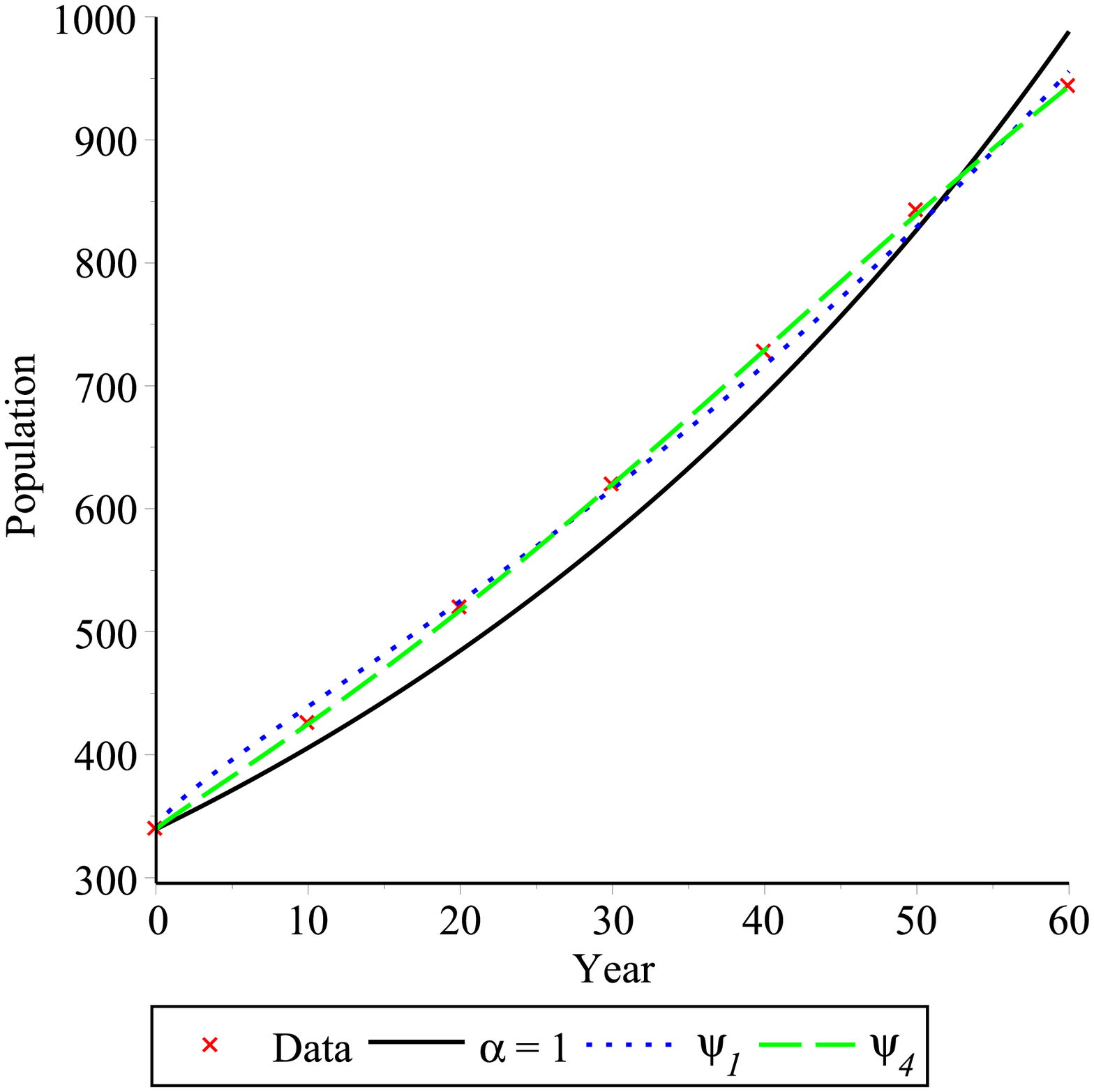}}
\subfigure[Asia.]{\includegraphics[width=6cm]{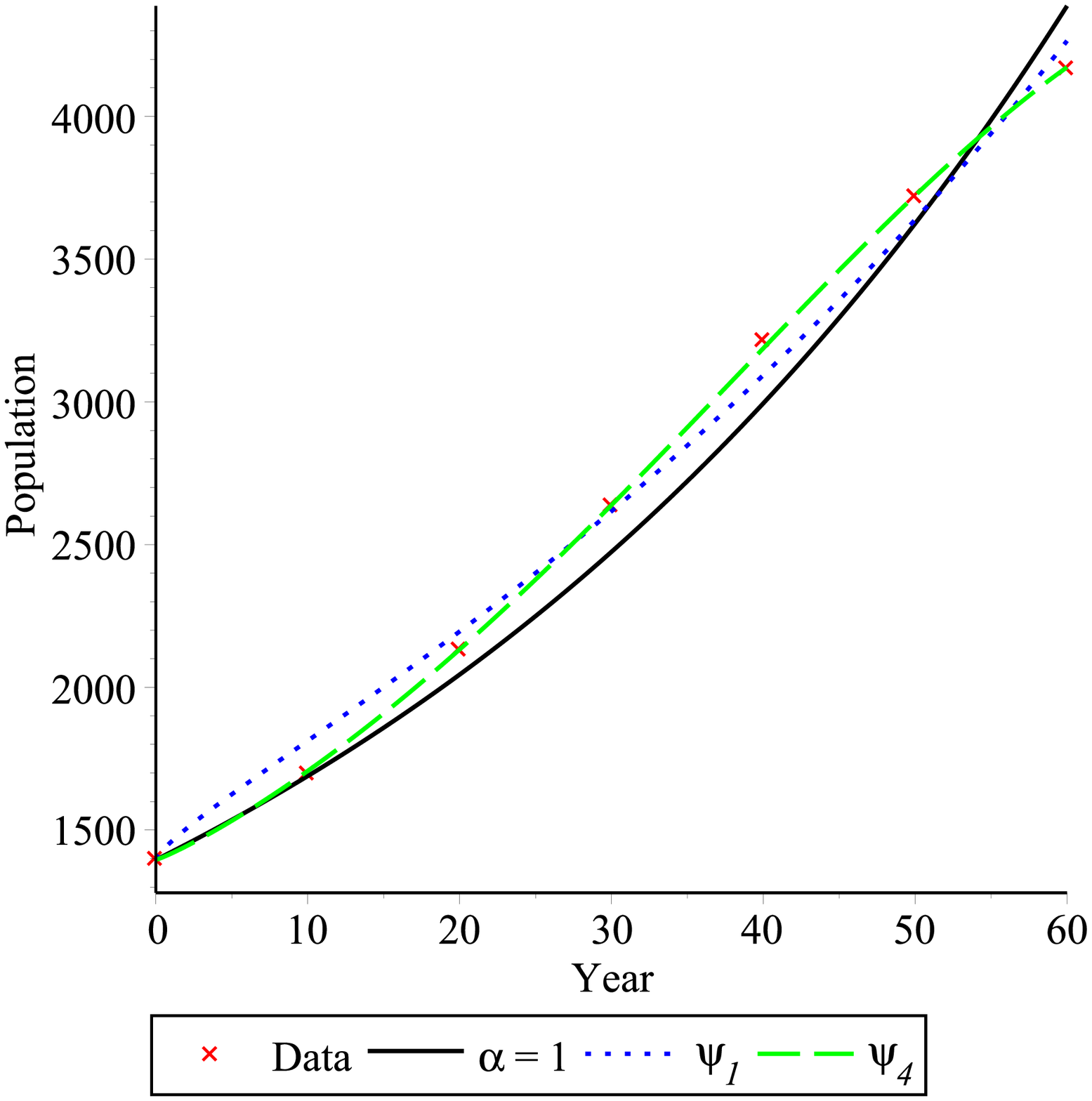}}
\subfigure[Europe.]{\includegraphics[width=6cm]{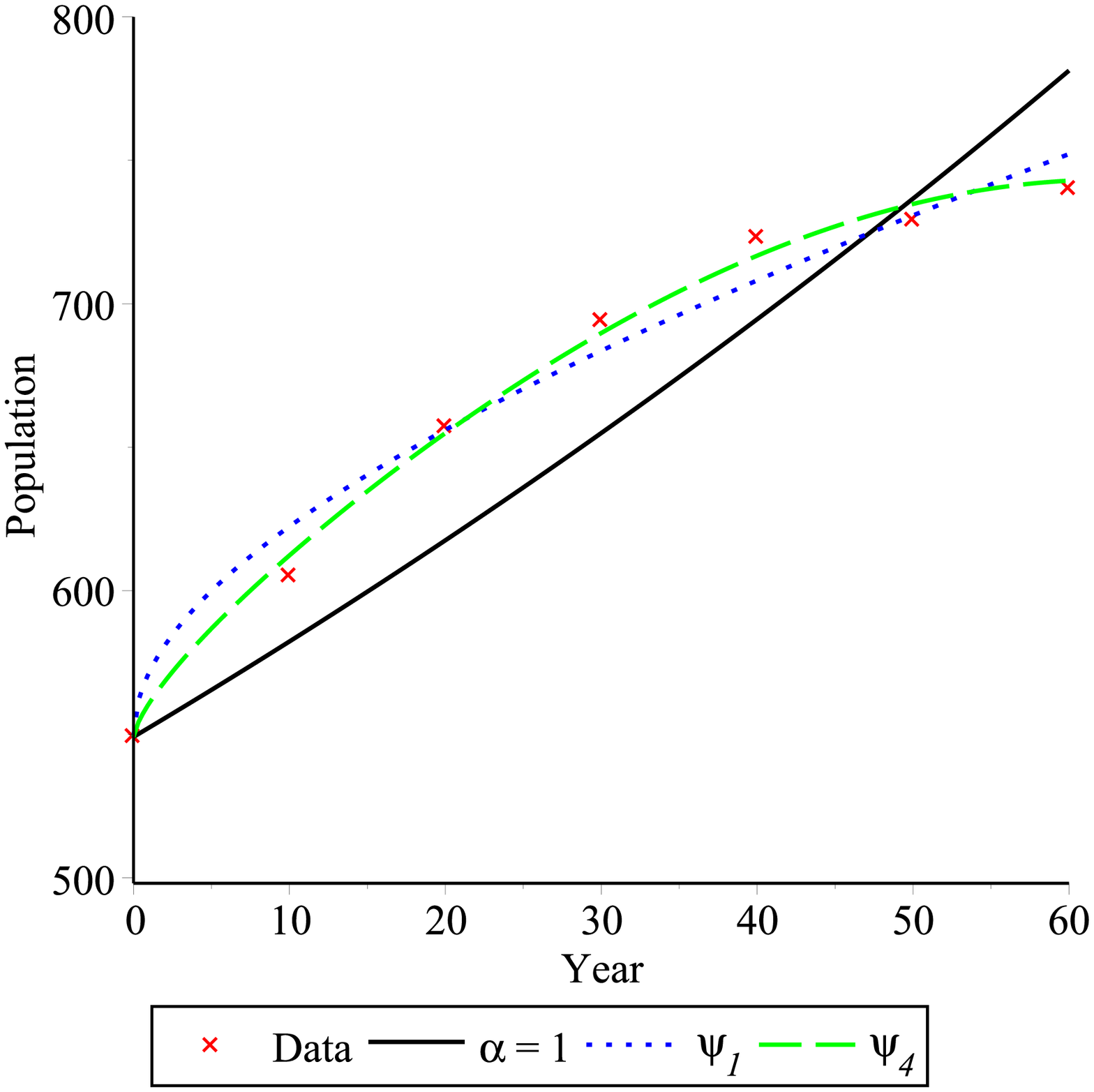}}
\caption{World population.}
\label{figPopulation}
\end{figure}

Historically, the most important law to characterize a population  growth is described by the differential equation
$$N'(t)=\lambda N(t),$$
where $\lambda$ is the population growth rate. The solution to this problem, with $N_0$ the initial number of individuals, is the exponential function
$$N(t)=N_0\exp(\lambda t).$$
When we consider the problem modeled by a FDE, that is, by the equation
$${^CD_{0+}^{\a,\p}} N(t)=\lambda N(t),$$
the solution is given by
$$N(t)=N_0E_\a(\lambda(\p(t)-\p(0))^\a).$$
We study the optimal curve that better fits with the data in four continents: Africa, America, Asia and Europe. For data source, we use the one available from the UN \cite{UN}. It consists of the number of individuals in the several continents, from 1950 until 2010, counted every 5 years. The initial value $t=0$ corresponds to the year 1950, with $N(t)$ measured in millions. The results are shown in Tables \ref{tab:Africa}, \ref{tab:America}, \ref{tab:Asia} and \ref{tab:Europe}.
{\small
 \begin{center}\begin{tabular}{|c|c|c|c|c|}
\hline
Model & $\lambda$ &  $\a$ & Error & Efficiency\\
\hline
Classical  & $0.025116$  & $1$ & $327.024434$ &\\
\hline
Fractional $\p_1$ & $0.022421$  &$ 1.03793$ &$ 236.643618$ &$27.637328\%$ \\
\hline
Fractional $\p_2$ ($b=0.7$) &$0.027381$ &$1.699578$  &$40.953972$ &$87.476785\%$ \\
\hline
\end{tabular}\captionof{table}{World population: Africa.}\label{tab:Africa}
\end{center}
\begin{center}\begin{tabular}{|c|c|c|c|c|}
\hline
Model & $\lambda$ &  $\a$ & Error & Efficiency\\
\hline
Classical  & $ 0.017806$  & $1$ & $6660.898290$ &\\
\hline
Fractional $\p_1$ & $ 0.042814$  &$0.723857$ &$685.697720$ &$89.705627\%$ \\
\hline
Fractional $\p_4$ ($b=60$) &$1.114633$ &$0 .915357$  &$10.987758$ &$99.835041\%$ \\
\hline
\end{tabular}\captionof{table}{World population: America.}\label{tab:America}
\end{center}
 \begin{center}\begin{tabular}{|c|c|c|c|c|}
\hline
Model & $\lambda$ &  $\a$ & Error & Efficiency\\
\hline
Classical  & $ 0.019087$  & $1$ & $ 1.378920\times 10^5$ &\\
\hline
Fractional $\p_1$ & $ 0.040394$  &$0.761743$ &$ 49095.37114$ &$64.395774\%$ \\
\hline
Fractional $\p_4$ ($b=50$) &$ 1.326809$ &$ 1.124085$  &$920.911794$ &$99.332150\%$ \\
\hline
\end{tabular}\captionof{table}{World population: Asia.}\label{tab:Asia}
\end{center}
 \begin{center}\begin{tabular}{|c|c|c|c|c|}
\hline
Model & $\lambda$ &  $\a$ & Error & Efficiency\\
\hline
Classical  & $0.005879$  & $1$ & $6246.195248$ &\\
\hline
Fractional $\p_1$ & $ 0.035060$  &$0.485031$ &$776.467956$ &$87.568945\%$ \\
\hline
Fractional $\p_4$  ($b=40$) &  $0.263725$ & $0.711280$  & $155.33140$ & $97.513184\%$ \\
\hline
\end{tabular}\captionof{table}{World population: Europe.}\label{tab:Europe}
\end{center}
}
For comparison, in Table \ref{tab:all}, we present the classical and the best fractional model, for the year 2015, with the estimated population in each of the continents analysed previously \cite{UN2}.
{\small
 \begin{center}\begin{tabular}{|c|c|c|c|c|c|}
\hline
Continent & Population &  Classical & Error & Fractional & Error\\
\hline
Africa  & $1166$  & $1171$ & $5$ & $1161$&$5$\\
\hline
America  & $991$  & $1080$ & $89$ & $990$&$1$\\
\hline
Asia  & $4384$  & $4826$ & $442$ & $4342$&$42$\\
\hline
Europe  & $743$  & $805$ & $62$ & $743$&$0$\\
\hline
\end{tabular}\captionof{table}{World population in 2015.}\label{tab:all}
\end{center}
}
In Figure \ref{figPopulation} we show the results.


\section*{Acknowledgments}

Work supported by Portuguese funds through the CIDMA - Center for Research and Development in Mathematics and Applications, and the Portuguese Foundation for Science and Technology (FCT-Funda\c{c}\~ao para a Ci\^encia e a Tecnologia), within project UID/MAT/04106/2013.


\end{document}